\documentclass[doublecol]{epl2} 

\usepackage{braket}
\usepackage[margin=1in]{geometry}
\usepackage{lipsum}
\usepackage{graphicx,caption}
\usepackage{hyperref}
\usepackage{bm}
\usepackage{amsmath}
\usepackage{amssymb}
\usepackage{braket}

\definecolor{massimiliano}{RGB}{0,0,255}

\def\ri{\bm{r}_i}

\def\dri{\dot{\textbf{r}}_i}
\def\ddri{\ddot{\textbf{r}}_i}

\def\r{\bm{r}}
\def\dr{\dot{\textbf{r}}}
\def\ddr{\ddot{\textbf{r}}}

\def\hex{\psi_{6i}}

\title{Entropy production of active Brownian particles going from liquid to hexatic and solid phases }
\shorttitle{Entropy production of active Brownian particles} 

\author{Massimiliano Semeraro \inst{1} \and Giuseppe Negro \inst{2} \and Antonio Suma \inst{1} \and Federico Corberi \inst{3} \and Giuseppe Gonnella \inst{1}}
\shortauthor{M. Semeraro, G. Negro, A. Suma, F. Corberi, G. Gonnella}

\institute{                    
  \inst{1} Dipartimento Interateneo di Fisica, Università degli Studi di Bari and INFN, Sezione di Bari, via Amendola 173, Bari I-70126, Italy\\
   \inst{2} School of Physics and Astronomy, University of Edinburgh, Peter Guthrie Tait Road, Edinburgh, EH9 3FD, UK\\
  \inst{3} Dipartimento di Fisica “E. R. Caianiello” and INFN, Gruppo Collegato di Salerno, via Giovanni Paolo II 132, 84084 Fisciano (SA), Italy \\
}

\abstract{Due to its inherent intertwinement with irreversibility, entropy production is a prime observable to monitor in systems of active particles. In this numerical study, entropy production in the liquid, hexatic and solid phases of a two-dimensional system of  active Brownian particles is examined at both average and fluctuation level. The trends of averages as functions of density show no singularity and marked changes in their derivatives at the hexatic-solid transition. Distributions show instead peculiar tail structures interpreted by looking at microscopic configurations. Particles in regions of low local order generate tail values according to different dynamical mechanisms: they move towards empty regions or bounce back and forth into close neighbours. The tail structures are reproduced by a simple single-particle model including an intermittent harmonic potential.}

\begin{document}

\maketitle

\section{Introduction}

In statistical mechanics, {\it entropy production} quantifies irreversibility of thermodynamical processes due to breaking of time-reversal symmetry: a positive value is a distinctive hallmark of non-equilibrium states and provides a measurement of their degree of irreversibility \cite{maes2003, andrieux2007, andrae2010, seifert2012, landi2013, peliti2021}. Furthermore, entropy production also proved as a valuable tool for characterizing non-equilibrium phase transitions \cite{shim2016, crosato2019}. As shown by numerous studies about master-equations \cite{noa2019, dasilva2020}, chemical \cite{nguyen2018, seara2021}, Ising-like \cite{zhang2016, barbosa2018, martynec2020}, lattice \cite{tome2012, barato2012, barbosa2018}, field \cite{suchanek2023} and active matter models \cite{nardini2017, negro2019, crosato2018, crosato2019, caballero2020, favuzzi2021, ferretti2022, ro2022, paoluzzi2024}, close to transition points or at the boundary between phase-separated states entropy production (or its derivative) peaks or even diverges. One can thus naturally wonder how general this behaviour is.

Investigation of average values but also of {\it fluctuations} of entropy production and related thermodynamical observables is central to fully characterize nonequilibrium systems. In this respect, {\it large deviations theory} \cite{dembo1984, denhollander2000} already had a deep impact in nonequilibrium statistical mechanics \cite{oono1989, ellis2007, lanford2007, touchette2009, cecconi2014, zamparo2023, semeraro2024}. A key concept here is the {\it large deviation principle}. For a quantity $\mathcal{S}_\tau$ integrated over a time $\tau$, this is satisfied if the asymptotics of its probability distribution $p(\mathcal{S}_\tau/\tau=s)$ can be described in terms of the {\it rate function} $I(s)\equiv\lim_{\tau\uparrow\infty}\ln(p(\mathcal{S}_\tau/\tau=s))/\tau$. Large deviations theory outperforms the central limit theorem as $I(s)$ assigns fluctuations of any amplitude an occurrence probability, and is additionally identified as a non-equilibrium analogue of free-energy.

The study of large deviations already proved successful in detecting peculiar fluctuation regimes. For a single Brownian particle, the rate function for the work injected by the thermal bath features a strictly convex central regime together with left and right linear stretches connected through second-order singular points \cite{farago2002, carollo2023b}. Similar results were found for single active particles and active work, i.e. particles driven by a self-propulsion force and work performed by the the latter \cite{semeraro2023a}. In both cases, linear stretches were associated to rare trajectories with large initial/final values, or {\it big jumps}. Singular behaviours for active work were also discovered in large systems of active particles \cite{cagnetta2017}. Here linear stretches were traced back to particles dragged by small cluster against their self-propelling direction. 

In this letter, we focus on Active Brownian Particles (ABPs) \cite{fily2012, romanczuk2012, bialke2013, redner2013, negro2022} in two dimensions and study how phase transitions at high density are reflected by the behaviour of entropy production at both average and fluctuation level. We also relate entropy production to the dynamics of topological defects characterizing these transitions. The phase diagram of ABPs has been carefully described in \cite{digregorio2018}. Here we consider intermediate activities for which at high enough densities a liquid-hexatic and an hexatic-solid phase transitions occur, with the hexatic and solid phases characterized by quasi-long-range orientational and translational order, respectively (see \autoref{fig:fig_1}$\textbf{(a)}$ and $\textbf{(b)}$). We find that the average entropy production as a function of density is not affected by any discontinuity, thus it cannot be employed as an order parameter. However, its derivative marks the hexatic-solid transition density with a pronounced change. Concerning fluctuations, in the liquid phase the log-probabilities result strictly convex, while in the ordered phases at intermediate times they show linear stretches or two-minima structures. We explain these features by looking at the trajectories of particles generating different entropy production contributions. Typically, particles locally arranged in hexatically ordered configurations are essentially locked and contribute to the convex part around the global minimum, while particles in regions of low local order or in proximity to vacancies result more mobile and contribute to the other branches. Finally, we propose a single-particle model able to reproduce these peculiar features by incorporating key dynamical aspects of the original one through an intermittent harmonic potential.

\section{Model}
We consider $N$ ABPs moving in a periodic square box of side $L$. Each ABP has unit diameter $\sigma_d$, self propels due to a force $\bm{a}_i=F_a\hat{\bm{n}}_i$ with constant amplitude $F_a$ and direction $\hat{\bm{n}}_i=(\cos\theta_i, \sin\theta_i)$ and has its dynamics ruled by the following Langevin equations
\begin{equation}
\begin{split}
    m\ddri&=-\gamma \dri+F_a\hat{\bm{n}}_i-\sum_{i\neq j}\nabla U_i(r_{ij})+\sqrt{2\gamma k_B T}~\bm{\xi}_i \\
    \dot{\theta}_i&=\sqrt{2D_\theta}~\eta_i
\end{split}~,
\label{eq:lang_eqs}
\end{equation}
where $i,j$ label particles, $m$ is their mass, $\ri$ is the position of the $i$-th particle, $\gamma$ is the friction coefficient, $k_B$ is the Boltzmann constant, $T$ is the environment temperature and $\bm{\xi}_i$ and $\bm{\eta}_i$ are two independent zero-mean white noises which satisfy $\braket{\bm{\xi}_i(t)\bm{\xi}_j(s)}=\delta_{ij}\delta(t-s)\bm{1}$ and $\braket{\eta_i(t)\eta_j(s)}=\delta_{ij}\delta(t-s)$. Particles interact via the shifted and truncated Mie potential $U(r)=4\epsilon [(\sigma_d/r)^{64}-(\sigma_d/r)^{32}]+\epsilon$ if $r<r_c=2^{1/32}\sigma_d$ and $0$ otherwise, with $\epsilon$ energy unit and $r_c$ location of the potential minimum. $D_\theta$ is a positive constant which is related to the self-propulsion persistence time as $\tau_p=1/D_\theta$, whence the persistence length $l_P=\tau_p F_a/\gamma$. As in \cite{das2018, semeraro2021}, we choose the convention $D_\theta=3 k_BT/(\sigma^2_d\gamma)$ which relates translational and rotational diffusion of spherical particle (see the Supplementary Material \cite{SM} for details). We consider three different system sizes, $N=128^2, 256^2$ and $512^2$, so as to address possible finite size effects. Moreover, in order to explore the ABP phase diagram from \cite{digregorio2018}, we introduce the density $\phi=N\pi\sigma^2_d/(4L^2)$, with $\phi_{lh},~\phi_{hs}$ and $\phi_{cp}\simeq 0.91$ denoting the liquid-hexatic and hexatic-solid transition densities and the close packing fraction, respectively, and the P\'eclet number $Pe=F_a\sigma_d/(k_BT)$, which compares the strength of the self-propulsion to thermal fluctuations. In the following we fix $m=1$, $\gamma=10$, $k_BT=0.05$ and $\epsilon=1$ so that the system is consistent with the overdamped picture and $\tau_p\simeq 66.67$. $\sigma_d,~m$ and $\epsilon$ serve as our length, mass and energy units \cite{allen2017}. For each system size, we explore the phase diagram region $10\leq Pe \leq 30$ without MIPS \cite{digregorio2018}. To monitor positional and orientational order, we introduce particle-wise the number of nearest neighbours $n_i$, which is extracted from a Voronoi tessellation, and the local hexatic order parameter $\hex$, which is defined as $\hex=\psi_6(\ri)\equiv\sum_{j=1}^{n_i}e^{\imath 6\theta_{ij}}/{n_i}$, where $\theta_{ij}$ is the angle formed by the segment connecting the centres of the $i$-th particle and the $j$-th out of $n_i$ neighbours and the $x$ axis. Details about numerical integration and sampling are reported in \cite{SM}.

\section{Entropy Production}
 In active particles systems, entropy production is an important subject \cite{fodor2016, mandal2017, pietzonka2017, grandpre2019, pietzonka2019, caprini2023} as parity of active force under time reversal transformation is not uniquely determined  \cite{puglisi2017, shankar2018, crosato2019, dabelow2019, caprini2019, fodor2022, byrne2022}. In principle, one could arbitrarily consider the latter either odd or even, as both choices are equally possible and supported by physical interpretations \cite{dabelow2019, fodor2022, byrne2022, oh2023}. Accordingly, two different expressions for the entropy production rate emerge,
\begin{equation}
    \dot{\tilde{\mathcal S}}^i_+\equiv\lim_{\tau\uparrow\infty}\frac{1}{\tau}\frac{F_a}{k_BT}\int_0^\tau \hat{\bm{n}}_i(s)\dri(s)~ds=\lim_{\tau\uparrow\infty}\dot{\tilde{s}}^i_+
    \label{eq:s_plus}
\end{equation}
\begin{equation}
    \dot{\tilde{\mathcal S}}^i_-\equiv\lim_{\tau\uparrow\infty}\frac{1}{\tau}\frac{F_a}{\gamma k_B T}\sum_{i\neq j}\int_0^\tau \hat{\bm{n}}_i(s)\nabla_i U(r_{ij})~ds=\lim_{\tau\uparrow\infty}\dot{\tilde{s}}^i_-~,
    \label{eq:s_minus}
\end{equation}
where the subscripts $\pm$ denote the active force is assumed even or odd (see \cite{SM} for details on vanishing boundary terms). Interestingly, these two expressions are not disconnected. One can in fact prove that $\braket{\dot{\tilde{\mathcal S}}^i_+}+\braket{\dot{\tilde{\mathcal S}}^i_-}=\braket{\dot{\tilde s}^i_+}+\braket{\dot{\tilde s}^i_-}=F_a^2/(\gamma k_B T)\equiv c_s$ particle-wise (see \cite{SM} for derivation). In the following we thus normalize the rates as $\dot{s}^i_\pm\equiv\dot{\tilde{s}}^i_\pm/c_s$ so that they become adimensional with range in $[0,1]$. We remark that $\dot{\tilde{\mathcal S}}^i_+$ is strictly related to the rate of active work, from which it differs only for a $1/T$ factor \cite{mandal2017, cagnetta2017, dabelow2019, pietzonka2019, nemoto2019, ekeh2020, keta2021}. We underline that active work is a fundamental observable in active systems as it captures the energy cost to sustain self propulsion \cite{dabelow2019, mandal2017, pietzonka2019, ekeh2020}, its distribution can be singular \cite{cagnetta2017, keta2021, semeraro2023a} and it defines the efficiency for active engines \cite{pietzonka2019, fodor2021}.

\section{Average Trends}
Firstly, note that $\dot{\mathcal S}_\pm^i$ sense particle aggregation. $\dot{\mathcal S}_+^i$ is large and positive when $\hat{\bm{n}}_i(s)\dri(s)>0$, i.e. when particles are driven by their activity unhindered by other particles, while it is large and negative when $\hat{\bm{n}}_i(s)\dri(s)<0$, i.e. when particles are dragged against their activity. $\dot{\mathcal S}_-^i$ is instead large and positive when $\hat{\bm{n}}_i(s)\nabla_i U(r_{ij})>0$, i.e. when particles are in close contact and push each other driven by activity.

\begin{figure}
       \includegraphics[width=1.0\columnwidth]{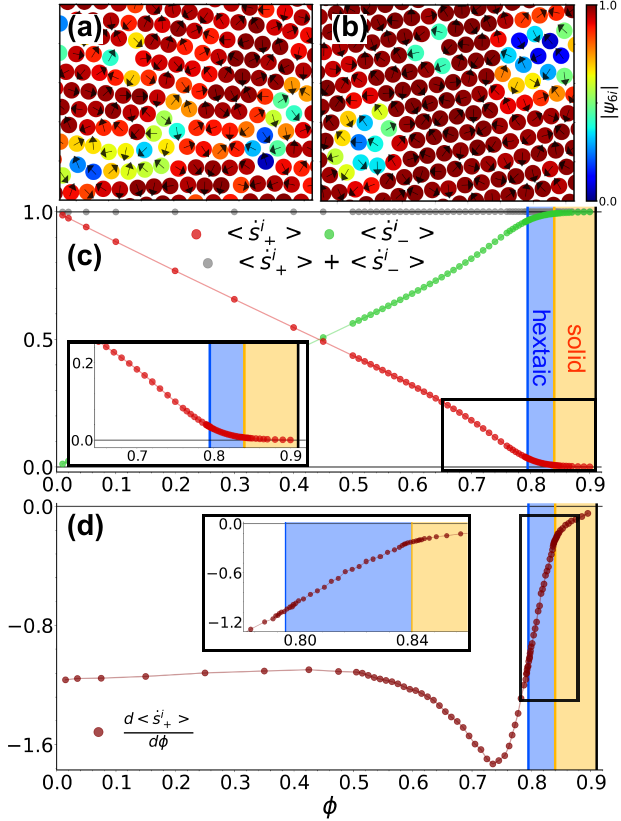}
\caption{\footnotesize{
$\textbf{(a)}$ and $\textbf{(b):}$ snapshots of a portion of the system at $\phi=0.82$ in the hexatic phase and $\phi=0.85$ in the solid one, respectively. Particles are coloured according to their $|\hex|$.
$\textbf{(c):}$ $\braket{\dot{s}^i_\pm}$ and $\braket{\dot{s}^i_+}+\braket{\dot{s}^i_-}$ as functions of $\phi$. The horizontal black lines report the limiting values $0,1$, the blue and yellow rectangles in the background highlight the regions of hexatic and solid order, with the two coloured vertical lines denoting the transition densities $\phi_{lh}\simeq 0.795$ and $\phi_{hs}\simeq 0.84$ \cite{digregorio2018} and the black one denoting the close packing $\phi_{cp}\simeq 0.91$. The inset reports an enlargement of the area enclosed into the rectangle.
$\textbf{(d):}$ $d\braket{\dot{s}^i_+}/d\phi$ as a function of $\phi$ with an enlargement of the area enclosed into the rectangle.
In all panels $Pe=10$ and $N=256^2$ are fixed.}}
\label{fig:fig_1}
\end{figure}

To make these remarks more quantitative, we look first at averages at $Pe=10$ and $N=256^2$. In \autoref{fig:fig_1}$\textbf{(c)}$ we report the trends of $\braket{\dot {s}^i_\pm}$ as functions of $\phi$ sampled over $\tau=10^3\gg\tau_p$. Let us focus on $\braket{\dot {s}^i_+}$ first. Note that, as $\phi$ is increased from $0$ to $\phi_{cp}\simeq 0.91$, particles have their mobility increasingly reduced and $\braket{\dot {s}^i_+}$ decreases from $1$ to $0$. In particular, $\braket{\dot { s}^i_+}$ decreases linearly up to $\phi\simeq 0.5$, i.e. well inside the dilute liquid phase. Notably, at larger $\phi$ $\braket{\dot {s}^i_+}$ seems to sense phase transitions. In fact, as shown by the inset in \autoref{fig:fig_1}$\textbf{(c)}$, as soon as particles start to order first orientationally at $\phi_{lh}\simeq 0.795$ and then positionally at $\phi_{hs}\simeq 0.84$ \cite{digregorio2018}, the decrease rate reduces until asymptotically zeroing at $\phi_{cp}$. This point is corroborated by the derivative reported \autoref{fig:fig_1}$\textbf{(d)}$. It is in fact almost constant up to $\phi=0.5$. At $\phi\sim 0.75$ it is interested by a bump which we attribute to the increased aggregation effect of activity at such densities. Reflecting the inset of \autoref{fig:fig_1}$\textbf{(c)}$, at larger densities the derivative is instead strictly increasing, and also shows a marked change around $\phi_{hs}$ which mirrors the passage from the hexatic phase, in which particles still are mobile, to the solid one, in which, apart from defects, they arrange orderly according to an hexagonal pattern (see \autoref{fig:fig_1}$\textbf{(a)}$ and $\textbf{(b)}$). We remark that our results are in agreement with \cite{caprini2023, caprini2023b}, where total entropy production rate for active solids is shown to reduce as $\phi$ is increased.

\autoref{fig:fig_1}$\textbf{(c)}$ shows that symmetric comments apply to $\braket{\dot {s}^i_-}$. In Figure S1 and S2 from \cite{SM} we confirm this scenario to hold for different $N$ and a larger $Pe$. We conclude remarking that, as shown in \autoref{fig:fig_1}$\textbf{(c)}$, the sum rule $\braket{\dot {s}^i_+}+\braket{\dot {s}^i_-}=1$ is satisfied at every $\phi$. We thus focus on $\dot{s}^i_+$.

\begin{figure*}[t!]
\center
       \includegraphics[width=2.0\columnwidth]{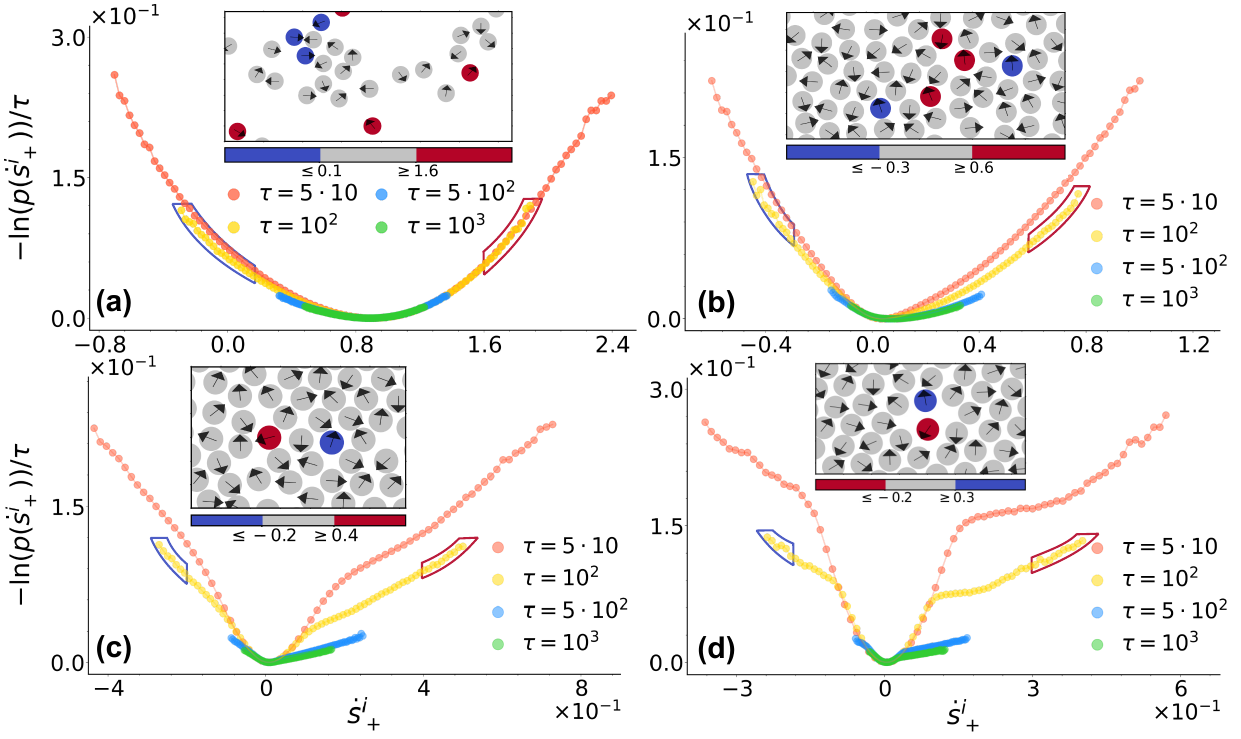}
\caption{\footnotesize{$\textbf{(a)}$ to $\textbf{(d):}$ curves $-\ln(p(\dot{ s}_+^i))/\tau$ at $\phi=0.1,~0.76,~0.82$ and $0.85$, respectively. For comparison, here and in the following these curves are shifted in such a way that their minimum is null. In each panel curves are coloured according to the $\tau$ during which $\dot{s}_+^i$ were sampled. Insets report snapshots of a portion of the system at corresponding $\phi$. As denoted by the horizontal bars, particles are coloured in red, grey and blue according to their $\dot{s}_+^i$ values sampled over $\tau=10^2\simeq\tau_p$. The left and right conditional intervals are highlighted in the mains by boxes with matching colours. Black arrows denote the instantaneous direction of the activity at the final sampling instant}}
\label{fig:fig_2}
\end{figure*}

\section{Fluctuations}
We now turn to the investigation of fluctuations, instrumental to better clarify the relation between entropy production and microscopic evolution. We sample $\dot{s}^i_+$ from all particles over increasing $\tau$ so as to estimate the curves $-\ln(p(\dot{s}^i_+))/\tau$, with $p(\dot{s}^i_+)$ probability distribution at corresponding $\tau$. In \autoref{fig:fig_2}$\textbf{(a)}$ to $\textbf{(d)}$ we report such curves at $\phi=0.1$ in a very dilute configuration and at $\phi=0.76,~0.82,~0.85$ well inside the liquid, hexatic and solid phases, respectively. In each case $\tau$ spans from $\tau=5\cdot 10 \simeq \tau_p\simeq$ to $\tau=10^3\gg \tau_p$. Looking first at $\textbf{(a)}$, we observe that the right branches of the curves collapse perfectly on each other, while the left branches do not, but rather tend to slightly flatten towards the horizontal axis. This is reminiscent of what found in \cite{cagnetta2017}, where at similar packing fractions but much larger $Pe$ the distribution of the active work rate is characterized by collapsing right branches and flattening left linear branches. In \cite{cagnetta2017}, this peculiar behaviour is ascribed to particles behaving in two different ways during sampling: some move freely without ever impacting other particles and then contributing to the right branches essentially as independent free particles, whereby collapse of the curves is expected; the remainder that contribute to the left branches are instead often dragged against their active forces by small clusters, hence increasing the probability of negative $\dot{s}^i_+$ values. Apart from particles bouncing into each other instead of being dragged due to our lower $Pe$, in our case the overall phenomenology is essentially the same. We therefore move to our novel results at larger $\phi$. \autoref{fig:fig_2}$\textbf{(b)}$ relative to $\phi=0.76$ shows that both branches of $-\ln(p(\dot{s}_+^i))/\tau$ have reduced convexity and tend to flatten. As for \autoref{fig:fig_2}$\textbf{(c)}$ and $\textbf{(d)}$ relative to $\phi=0.82,0.85$, we instead observe a net distinction between a central parabolic regime around $\dot{s}^i_+\simeq \braket{\dot{ s}^i_+}\simeq 0$ and left and right branches with peculiar shapes, the latter attaching to the parabolic region at decreasing $\dot{s}^i_+$ as $\tau$ is increased. An intuition about these peculiar shapes is provided by the insets of \autoref{fig:fig_2}, which report snapshots with particles coloured according to their $\dot{ s}^i_+$ values sampled over $\tau=10^2\simeq 1.5\tau_p$ (yellow points in the mains). As apparent from the inset of \autoref{fig:fig_2}$\textbf{(d)}$, we observe that $\dot{s}^i_+$ values close to $\braket{\dot{s}^i_+}$ and far in the tails seem to be associated to particles trapped in regions of high local positional order and with more freedom of movement in regions of low local one.

\begin{figure*}[t!]
\center
       \includegraphics[width=1.95\columnwidth]{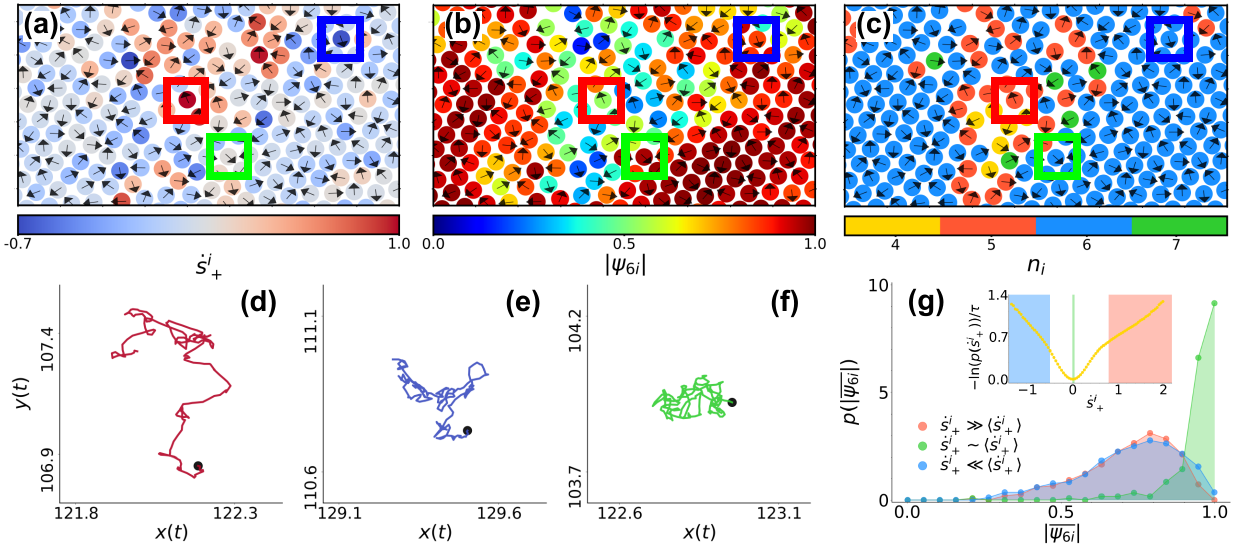}
\caption{\footnotesize{$\textbf{(a)}$ to $\textbf{(c):}$ snapshots of a portion of the system at $\phi=0.82$. Particles are coloured according to their $\dot{s}^i_+,~|\psi_{6i}|$ and $n_i$ values, the former sampled over $\tau=10$, the latter two extracted at the final sampling instant. Rectangles are instrumental to the bottom panels.
$\textbf{(d)}$ to $\textbf{(f):}$ sample trajectories relative to the centre of particles with $\dot{s}_+^i\simeq 1.04, -0.76$ and $8.74\cdot 10^{-3}$, the first two far in the tails, the latter $\sim \braket{\dot{s}_+^i}\sim 1.22\cdot 10^{-2}$ (see inset of $\textbf{(g)}$). Black dots denote the trajectories initial points, while their colours matches the rectangle from $\textbf{(a)}$ to $\textbf{(c)}$ individuating the particles they belong to. For comparison, in all panels the $x$ and $y$ axes span the same interval length $0.7\sigma_d$.
$\textbf{(g):}$ distributions of $|\bar\psi_{6i}|$ conditioned on $\dot{s}^i_+\simeq\braket{\dot{s}^i_+}\sim 1.22\cdot 10^{-2}$, $\dot{s}^i_+\geq0.8\gg \braket{\dot{s}^i_+}$ and $\dot{s}^i_+\leq-0.5\ll \braket{\dot{s}^i_+}$. The inset reports the curve $-\ln(p(\dot{s}_+^i))/\tau$ for $\tau=10$ built using $\dot{s}_+^i$ values from all particles. Rectangles in the background are coloured according to the legend from the main and highlight the conditioning intervals.}}
\label{fig:fig_3}
\end{figure*}

In order to quantitatively support this intuition, we now relate $\dot{s}^i_+$ to $|\psi_{6i}|$, $n_i$ values. In \autoref{fig:fig_3}$\textbf{(a)}$ to $\textbf{(c)}$ we report snapshots of the system at $\phi=0.82$ with particles respectively coloured according to their $\dot{s}^i_+$, $|\psi_{6i}|$ and $n_i$ values (see Figure S3 and S5 in \cite{SM} for similar figures at $\phi=0.76$ and $0.85$). To better access $\dot{s}^i_+$ values far in the tails, here we reduce the sampling interval to $\tau=10<\tau_p$. $\psi_{6i},~n_i$ are instead extracted at the final sampling instant. Interestingly, a quick comparison between panels reveals that particles with $\dot{s}^i_+\sim\braket{\dot{s}^i_+}$ are often characterized by final $|\psi_{6i}|\sim 1,~n_i=6$ values, while particles with $|\dot{s^i_+}|\gg\braket{\dot{S^i_+}}$ are associated to $\psi_{6i}<1,~n_i<6$ values. Hence the former are hexatically ordered, while the latter are located close or even into vacancies at the final time. Complementary information is provided by the sample trajectories from \autoref{fig:fig_3}$\textbf{(d)}$ to $\textbf{(f)}$ associated to particles from the upper panels highlighted by matching rectangles: $\textbf{(d)}$ shows that large positive values are due to particles moving towards regions where free space is available, so that velocity and activity are often parallel; the irregular trajectory from $\textbf{(e)}$ remarks instead that negative values are associated to particles still moving in freer regions, but often bouncing into or pushed by close neighbours, so that now velocity and activity are often anti-parallel; finally $\textbf{(f)}$ shows that values close to average are due to particles which remain essentially locked due to the trapping exerted by close neighbours. A more direct visualization of the mechanisms just described is offered by Movie S1, S2 and S3 from \cite{SM}. The consistency of this scenario during the entire time interval is instead proved in \autoref{fig:fig_3}$\textbf{(g)}$. Here we report the distribution of $|\bar\psi_{6i}|$, i.e. of the modulus of $\psi_{6i}$ time-averaged over $\tau$, conditioned on the intervals $\dot{s}^i_+\gg \braket{\dot{s}^i_+}$, $\dot{s}^i_+\ll \braket{\dot{s}^i_+}$ and $\dot{s}^i_+\simeq\braket{\dot{s}^i_+}$ highlighted in the inset. It is apparent that for particles with $\dot{s}^i_+\simeq\braket{\dot{s}^i_+}$, $|\bar\psi_{6i}|\sim 1$, i.e. these particles remain essentially locally hexatically ordered over the entire sampling interval. On the contrary, particles associated to high and low $\dot{s}^i_+$ values are characterized by $|\bar\psi_{6i}|<1$ values, i.e. they are located in regions of low local hexatic order. Figure S6 from \cite{SM} shows that an analysis of $\bar n_i$ leads to coherent and complementary results. 

\section{A phenomenological model}
Having grasped the essential dynamical aspects generating peculiar distribution shapes, we now introduce a simple single-particle model incorporating the former and reproducing the latter. The basic idea is the following: when in regions of high (low) local order, particles are considered effectively trapped (free). The simple phenomenological model we therefore employ is inspired by Brownian resetting studies and includes a single ABP under the action of an intermittent harmonic potential \cite{besga2020, santra2021, mercado2022}: 
\begin{equation}
\begin{split}
    m\ddr&=-\gamma \dr+F_a\hat{\bm{n}}-k\lambda(t)\r(t)+\sqrt{2\gamma k_B T}~\bm{\xi} \\
    \dot{\theta}&=\sqrt{2D_\theta}~\eta
\end{split}~.
\label{eq:lang_eqs_int}
\end{equation}
Here $k$ is the elastic constant of a harmonic potential $U(\r)=k\r^2/2$ and $\lambda(t)$ is a dichotomous noise alternatively taking values $0$ and $1$, while all other symbols keep the same meaning as in \autoref{eq:lang_eqs}. The time intervals $\tilde \tau$ during which $\lambda(t)$ is $0$ or $1$ are extracted from the exponential distributions $p_{f/h}(\tilde\tau)=e^{-{\tilde\tau}/\tau_{f/h}}/\tau_{f/h},\tilde\tau\geq0$, where the subscript $f$ $(h)$ refers to $\lambda(t)=0$ ($1$). The average fraction of times during which the particle is free or confined thus are $\zeta_{f/h}=\tau_{f/h}/(\tau_f+\tau_h)$. We fix $\tau_f=\tau_p$ and set $\tau_h=\alpha\cdot\tau_f,~\alpha >0$. The first equality ensures that the ABP is on average free for intervals $\tau_p$ during which at $Pe\sim10$ it travels distances $l_p\sim3\sim\sigma_d$ comparable to the ones usually travelled by particles in regions of low local order (see \autoref{fig:fig_1}$\textbf{(a)}$ and $\textbf{(b)}$). The second one instead simply tunes $\tau_h$ with $\tau_f$ as reference: with $\alpha\gg1$ ($\ll1$) the particle is mostly trapped (free) like particles in regions of high (low) local order. We remark that once $\tau_h/\tau_f=\alpha$ is fixed, choosing a different $\tau_f$ only results in a rescaling of the times at which same log-curves are observed.

\begin{figure}
       \includegraphics[width=0.95\columnwidth]{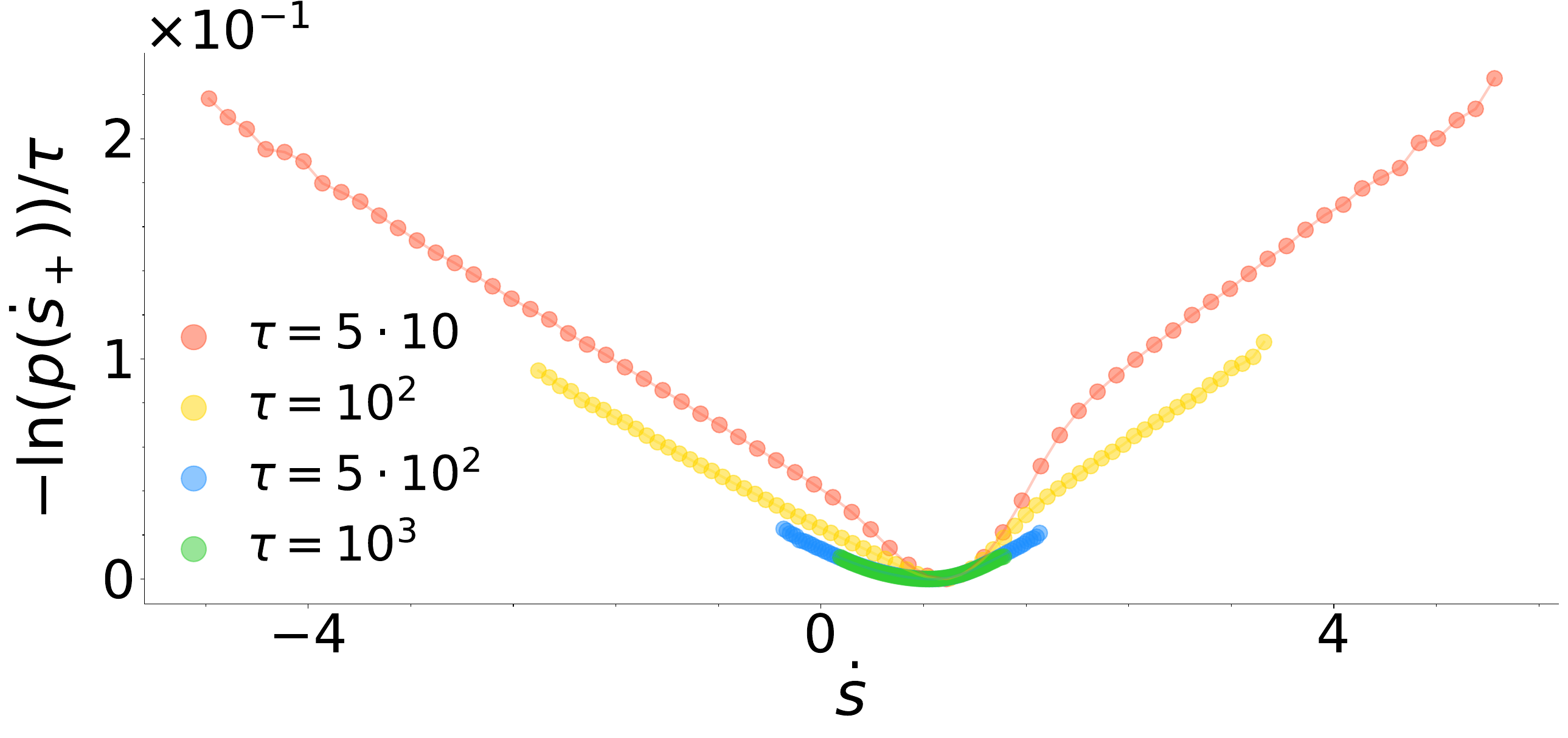}
\caption{\footnotesize{Curves $-\ln(p(\dot{s}_+))/\tau$ at $Pe=10,~k=1,~\alpha=0.1$ for different sampling times $\tau$ in the single-particle model \autoref{eq:lang_eqs_int}.}}
\label{fig:fig_4}
\end{figure}

We now provide proof that our simple model actually works. To reproduce results from the hexatic phase, in which particles are often far and softly impact each other, we set $Pe=10,k=1$ and $\alpha=0.1$, so that $\zeta_h\sim 9\%$. The resulting curves $-\ln(p(\dot{s}_+)/\tau$ for our single ABP from \autoref{fig:fig_4} clearly resemble the ones from \autoref{fig:fig_2}$\textbf{(c)}$, with $\dot{s}_+$ values far in the tails and $\sim\braket{\dot{s}_+}$ due to particles trapped on average $50\%\gg\zeta_h$ and $\% 5\sim \zeta_h$ of the time, respectively. As shown in Figure S7 from \cite{SM}, curves similar to the ones from \autoref{fig:fig_2}$\textbf{(b)}$ and $\textbf{(d)}$ relative to the liquid and solid phases can be obtained by fine tuning $\alpha$. In general, as shown by the trajectories in Fig. S7 high-$\dot{s}_+$ particles persistently move pushed by the active force while free, so that velocity and active force are often parallel. Instead, low-$\dot{s}_+$ particles are often trapped when far away from the centre of the potential, so that while being pushed towards it their velocities and active forces are often anti-parallel. These mechanisms thus reproduce the motion of particles in regions of low local order which either move towards empty space or bounce into neighbours. 

\section{Conclusions}
In this letter, we numerically studied entropy production rates for large systems of ABPs both at the average and fluctuation level. We explored regions of the phase diagram in which MIPS does not occur, but rather the system transitions from a liquid to a hexatic and a solid phase upon increasing density. Concerning averages, we found their trends as functions of density to monotonically vary according to the degree of mobility of particles and closeness to neighbours without discontinuity, thus denying them the role of order parameters. Their derivatives mark instead a pronounced change around the hexatic-solid transition. Concerning fluctuations, at intermediate times we detected peculiar tail structures in distributions which we explained by looking at microscopic configurations. We found that values far in the tails are associated to particles in regions of low local order where they can either move towards empty regions or bounce into other particles. We proved the connection between tail values and lost local order by measuring particle-wise the local hexatic order parameter and neighbour number. We also introduced a simple single-particle model able to reproduce these peculiar distribution shapes. This is based on the idea that when in regions of high (low) local order, particles are effectively trapped (free), and therefore includes an intermittent harmonic potential. In the future we plan to perturb our system with an asymmetric ratchets potential \cite{mcdermott2016, derivaux2022, semeraro2023c} to study its effect on entropy fluctuations and thermodynamical efficiency of the resulting device. It would also be interesting to extend our analysis to dynamical transitions occurring in active matter models described through orientational order parameters \cite{head2024}.

\acknowledgments

\noindent 
Access to Bari ReCaS e-Infrastructure funded through PON Research and Competitiveness 2007-2013 Call 254 Action and to high-performance computing resources granted by the CINECA award No. ISCRA IsCb1 AcT made this work possible. M.S., G.N., A.S., G.G. acknowledge financial support by MUR projects PRIN 2020/PFCXPE, PRIN 2022/HNW5YL and Quantum Sensing and Modelling for One-Health (QuaSiModO). F.C. acknowledges financial support by MUR PRIN 2022 PNRR.

\bibliographystyle{eplbib.bst} 
\bibliography{Refs.bib}

\begin{thebibliography}{10}
\expandafter\ifx\csname url\endcsname\relax\def\url#1{\texttt{#1}}\fi

\bibitem{maes2003}
\Name{Maes C. \and Neto{\v{c}}n{\`y} K.} \REVIEW{J. stat.
  phys.}{110}{2003}{269}.

\bibitem{andrieux2007}
\Name{Andrieux D., Gaspard P., Ciliberto S., Garnier N., Joubaud S. \and
  Petrosyan A.} \REVIEW{Phys. Rev. Lett.}{98}{2007}{150601}.

\bibitem{andrae2010}
\Name{Andrae B., Cremer J., Reichenbach T. \and Frey E.} \REVIEW{Phys. Rev.
  Lett.}{104}{2010}{218102}.

\bibitem{seifert2012}
\Name{Seifert U.} \REVIEW{Rep. Prog. Phys.}{75}{2012}{126001}.

\bibitem{landi2013}
\Name{Landi G.~T., Tomé T. \and de~Oliveira M.~J.} \REVIEW{J. Phys.
  A}{46}{2013}{395001}.

\bibitem{peliti2021}
\Name{Peliti L. \and Pigolotti S.} \Book{Stochastic {T}hermodynamics: {A}n
  {I}ntroduction} (Princeton) 2021.

\bibitem{shim2016}
\Name{Shim P.-S., Chun H.-M. \and Noh J.~D.} \REVIEW{Phys. Rev.
  E}{93}{2016}{012113}.

\bibitem{crosato2019}
\Name{Crosato E., Prokopenko M. \and Spinney R.~E.} \REVIEW{Phys. Rev.
  E}{100}{2019}{042613}.

\bibitem{noa2019}
\Name{Noa C. E.~F., Harunari P.~E., de~Oliveira M.~J. \and Fiore C.~E.}
  \REVIEW{Phys. Rev. E}{100}{2019}{012104}.

\bibitem{dasilva2020}
\Name{da~Silva R., de~Oliveira M.~J., Tom\'e T. \and Drugowich~de Fel\'{\i}cio
  J.~R.} \REVIEW{Phys. Rev. E}{101}{2020}{012130}.

\bibitem{nguyen2018}
\Name{Nguyen B., Seifert U. \and Barato A.~C.} \REVIEW{J. Chem.
  Phys.}{149}{2018}{045101}.

\bibitem{seara2021}
\Name{Seara D.~S., Machta B.~B. \and Murrell M.~P.} \REVIEW{Nat.
  Comm.}{12}{2021}{392}.

\bibitem{zhang2016}
\Name{Zhang Y. \and Barato A.~C.} \REVIEW{J. Stat. Mech.}{2016}{2016}{113207}.

\bibitem{barbosa2018}
\Name{Barbosa O.~A. \and Tomé T.} \REVIEW{J. Stat. Mech}{2018}{2018}{063202}.

\bibitem{martynec2020}
\Name{Martynec T., Klapp S. H.~L. \and Loos S. A.~M.} \REVIEW{New J.
  Phys.}{22}{2020}{093069}.

\bibitem{tome2012}
\Name{Tom\'e T. \and de~Oliveira M.~J.} \REVIEW{Phys. Rev.
  Lett.}{108}{2012}{020601}.

\bibitem{barato2012}
\Name{Barato A.~C. \and Hinrichsen H.} \REVIEW{J. Phys. A}{45}{2012}{115005}.

\bibitem{suchanek2023}
\Name{Suchanek T., Kroy K. \and Loos S. A.~M.} \REVIEW{Phys. Rev.
  Lett.}{131}{2023}{258302}.

\bibitem{nardini2017}
\Name{Nardini C., Fodor E., Tjhung E., van Wijland F., Tailleur J. \and Cates
  M.~E.} \REVIEW{Phys. Rev. X}{7}{2017}{021007}.

\bibitem{negro2019}
\Name{Negro G., Carenza L.~N., Lamura A., Tiribocchi A. \and Gonnella G.}
  \REVIEW{Soft Matter}{15}{2019}{8251}.

\bibitem{crosato2018}
\Name{Crosato E., Spinney R.~E., Nigmatullin R., Lizier J.~T. \and Prokopenko
  M.} \REVIEW{Phys. Rev. E}{97}{2018}{012120}.

\bibitem{caballero2020}
\Name{Caballero F. \and Cates M.~E.} \REVIEW{Phys. Rev.
  Lett.}{124}{2020}{240604}.

\bibitem{favuzzi2021}
\Name{Favuzzi I., Carenza L.~N., Corberi F., Gonnella G., Lamura A. \and Negro
  G.} \REVIEW{Soft Materials}{19}{2021}{334}.

\bibitem{ferretti2022}
\Name{Ferretti F., Grosse-Holz S., Holmes C., Shivers J.~L., Giardina I., Mora
  T. \and Walczak A.~M.} \REVIEW{Phys. Rev. E}{106}{2022}{034608}.

\bibitem{ro2022}
\Name{Ro S., Guo B., Shih A., Phan T.~V., Austin R.~H., Levine D., Chaikin
  P.~M. \and Martiniani S.} \REVIEW{Phys. Rev. Lett.}{129}{2022}{220601}.

\bibitem{paoluzzi2024}
\Name{Paoluzzi M., Levis D., Crisanti A. \and Pagonabarraga I.}
  \REVIEW{arXiv:2310.03423}{}{2024}{}.

\bibitem{dembo1984}
\Name{Dembo A. \and Zeitouni O.} \Book{Large {D}eviations {Te}chniques and
  {A}pplications} 2nd Edition (Springer - New York) 1988.

\bibitem{denhollander2000}
\Name{den Hollander F.} \Book{Large {D}eviations} (AMS) 2000.

\bibitem{oono1989}
\Name{Oono Y.} \REVIEW{Prog. Th. Phy Suppl.}{99}{1989}{165}.

\bibitem{ellis2007}
\Name{Ellis R.~S.} \Book{Entropy, large deviations, and statistical mechanics}
  (Springer) 2007.

\bibitem{lanford2007}
\Name{Lanford O.~E.} \Book{Entropy and equilibrium states in classical
  statistical mechanics} in \Book{Stat. mech. math. prob.} (Springer) 2007 pp.
  1--113.

\bibitem{touchette2009}
\Name{Touchette H.} \REVIEW{Phys. Rep.}{478}{2009}{1}.

\bibitem{cecconi2014}
\Name{Cecconi F., Cencini M., Puglisi A., Vergni D. \and Vulpiani A.}
  \Book{From the law of large numbers to large deviation theory in statistical
  physics: {A}n introduction} (Springer) 2014.

\bibitem{zamparo2023}
\Name{Zamparo M. \and Semeraro M.} \REVIEW{J. Math. Phys.}{64}{2023}{}.

\bibitem{semeraro2024}
\Name{Semeraro M., Suma A. \and Negro G.} \REVIEW{Entropy}{26}{2024}{}.

\bibitem{farago2002}
\Name{Farago J.} \REVIEW{J. stat. phys.}{107}{2002}{781}.

\bibitem{carollo2023b}
\Name{Carollo G.~B., Semeraro M., Gonnella G. \and Zamparo M.} \REVIEW{J. Phys.
  A}{56}{2023}{435003}.

\bibitem{semeraro2023a}
\Name{Semeraro M., Gonnella G., Suma A. \and Zamparo M.} \REVIEW{Phys. Rev.
  Lett.}{131}{2023}{158302}.

\bibitem{cagnetta2017}
\Name{Cagnetta F., Corberi F., Gonnella G. \and Suma A.} \REVIEW{Phys. Rev.
  Lett.}{119}{2017}{158002}.

\bibitem{fily2012}
\Name{Fily Y. \and Marchetti M.~C.} \REVIEW{Phys. Rev.
  Lett.}{108}{2012}{235702}.

\bibitem{romanczuk2012}
\Name{Romanczuk P., B{\"a}r M., Ebeling W., Lindner B. \and Schimansky-Geier
  L.} \REVIEW{EPJ Spec. Top.}{202}{2012}{1}.

\bibitem{bialke2013}
\Name{Bialké J., Löwen H. \and Speck T.} \REVIEW{EPL}{103}{2013}{30008}.

\bibitem{redner2013}
\Name{Redner G.~S., Hagan M.~F. \and Baskaran A.} \REVIEW{Phys. Rev.
  Lett.}{110}{2013}{055701}.

\bibitem{negro2022}
\Name{{Negro, G.}, {Caporusso, C. B.}, {Digregorio, P.}, {Gonnella, G.},
  {Lamura, A.} \and {Suma, A.}} \REVIEW{Eur. Phys. J. E}{45}{2022}{75}.

\bibitem{digregorio2018}
\Name{Digregorio P., Levis D., Suma A., Cugliandolo L.~F., Gonnella G. \and
  Pagonabarraga I.} \REVIEW{Phys. Rev. Lett.}{121}{2018}{098003}.

\bibitem{das2018}
\Name{Das S., Gompper G. \and Winkler R.~G.} \REVIEW{New J.
  Phys.}{20}{2018}{015001}.

\bibitem{semeraro2021}
\Name{Semeraro M., Suma A., Petrelli I., Cagnetta F. \and Gonnella G.}
  \REVIEW{J. Stat. Mech.}{2021}{2021}{123202}.

\bibitem{SM}
\Name{Semeraro M., Negro G., Suma A., Corberi F. \and Gonnella G.}
  \Book{Supplementary {M}aterial}.

\bibitem{allen2017}
\Name{Allen M.~P. \and Tildesley D.~J.} \Book{Computer simulation of liquids}
  (Oxford) 2017.

\bibitem{fodor2016}
\Name{Fodor E., Nardini C., Cates M.~E., Tailleur J., Visco P. \and van Wijland
  F.} \REVIEW{Phys. Rev. Lett.}{117}{2016}{038103}.

\bibitem{mandal2017}
\Name{Mandal D., Klymko K. \and DeWeese M.~R.} \REVIEW{Phys. Rev.
  Lett.}{119}{2017}{258001}.

\bibitem{pietzonka2017}
\Name{Pietzonka P. \and Seifert U.} \REVIEW{J. Phys. A}{51}{2017}{01LT01}.

\bibitem{grandpre2019}
\Name{GrandPre T., Klymko K., Mandadapu K.~K. \and Limmer D.~T.} \REVIEW{Phys.
  Rev. E}{103}{2021}{012613}.

\bibitem{pietzonka2019}
\Name{Pietzonka P., Fodor E., Lohrmann C., Cates M.~E. \and Seifert U.}
  \REVIEW{Phys. Rev. X}{9}{2019}{041032}.

\bibitem{caprini2023}
\Name{Caprini L., Marini Bettolo~Marconi U., Puglisi A. \and Löwen H.}
  \REVIEW{J. Chem. Phys.}{159}{2023}{041102}.

\bibitem{puglisi2017}
\Name{Puglisi A. \and Marini Bettolo~Marconi U.} \REVIEW{Entropy}{19}{2017}{}.

\bibitem{shankar2018}
\Name{Shankar S. \and Marchetti M.~C.} \REVIEW{Phys. Rev. E}{98}{2018}{020604}.

\bibitem{dabelow2019}
\Name{Dabelow L., Bo S. \and Eichhorn R.} \REVIEW{Phys. Rev.
  X}{9}{2019}{021009}.

\bibitem{caprini2019}
\Name{Caprini L., Marconi U. M.~B., Puglisi A. \and Vulpiani A.} \REVIEW{J.
  Stat. Mech.}{2019}{2019}{053203}.

\bibitem{fodor2022}
\Name{Fodor E., Jack R.~L. \and Cates M.~E.} \REVIEW{Ann. Rev. Cond. Matt.
  Phys.}{13}{2022}{}.

\bibitem{byrne2022}
\Name{O’Byrne J., Kafri Y., Tailleur J. \and van Wijland F.} \REVIEW{Nat.
  Rev. Phys.}{4}{2022}{167183}.

\bibitem{oh2023}
\Name{Oh Y. \and Baek Y.} \REVIEW{Phys. Rev. E}{108}{2023}{024602}.

\bibitem{nemoto2019}
\Name{Nemoto T., Fodor {\'E}., Cates M.~E., Jack R.~L. \and Tailleur J.}
  \REVIEW{Phys. Rev. E}{99}{2019}{022605}.

\bibitem{ekeh2020}
\Name{Ekeh T., Cates M.~E. \and Fodor E.} \REVIEW{Phys. Rev.
  E}{102}{2020}{010101}.

\bibitem{keta2021}
\Name{Keta Y.-E., Fodor {\'E}., van Wijland F., Cates M.~E. \and Jack R.~L.}
  \REVIEW{Phys. Rev. E}{103}{2021}{022603}.

\bibitem{fodor2021}
\Name{Fodor E. \and Cates M.~E.} \REVIEW{EPL}{134}{2021}{10003}.

\bibitem{caprini2023b}
\Name{Caprini L., Löwen H. \and Marconi U. M.~B.} \REVIEW{J. Phys.
  A}{56}{2023}{465001}.

\bibitem{besga2020}
\Name{Besga B., Bovon A., Petrosyan A., Majumdar S.~N. \and Ciliberto S.}
  \REVIEW{Phys. Rev. Res.}{2}{2020}{032029}.

\bibitem{santra2021}
\Name{Santra I., Das S. \and Nath S.~K.} \REVIEW{J. Phys. A}{54}{2021}{334001}.

\bibitem{mercado2022}
\Name{Mercado-Vásquez G., Boyer D. \and Majumdar S.~N.} \REVIEW{J. Stat.
  Mech.}{2022}{2022}{093202}.

\bibitem{mcdermott2016}
\Name{McDermott D., Reichhardt C. J.~O. \and Reichhardt C.} \REVIEW{Soft
  Matter}{12}{2016}{8606}.

\bibitem{derivaux2022}
\Name{Derivaux J.-F., Jack R.~L. \and Cates M.~E.} \REVIEW{J. Stat.
  Mech.}{2022}{2022}{043203}.

\bibitem{semeraro2023c}
\Name{Semeraro M., Gonnella G., Lippiello E. \and Sarracino A.}
  \REVIEW{Symmetry}{15}{2023}{200}.

\bibitem{head2024}
\Name{Head L.~C., Dor{\'e} C., Keogh R.~R., Bonn L., Negro G., Marenduzzo D.,
  Doostmohammadi A., Thijssen K., L{\'o}pez-Le{\'o}n T. \and Shendruk T.~N.}
  \REVIEW{Nat. Phys.}{20}{2024}{492}.

\end{thebibliography}

\end{document}